\documentclass[preprint,letterpaper,prl,aps,amsmath,amssymb,groupedaddress,showpacs]{revtex4}
\usepackage{graphicx}

\def \oli 	{LiFePO$_4$}
\def \fepo  	{FePO$_4$}
\def \olix 	{Li$_x$FePO$_4$}
\def \ldapu 	{LDA$+ U$}

\begin{document}

\title{Phase Separation in Li$_x$FePO$_4$ Induced by Correlation Effects}

\author{F. Zhou}
\affiliation{Department of Physics,
 Massachusetts Institute of Technology,
 Cambridge, 02139 MA, USA}
\email{fzhou@mit.edu}
\author{C. A. Marianetti, M. Cococcioni, D. Morgan, G. Ceder}
\email{gceder@mit.edu}
\affiliation{Department of Material Science and Engineering,
 Massachusetts Institute of Technology,
 Cambridge, 02139 MA, USA}

\date{\today}

\begin{abstract}
We report on a significant failure of the local density approximation (LDA) and the 
generalized gradient approximation (GGA) to reproduce the phase stability and 
thermodynamics of mixed-valence Li$_x$FePO$_4$ compounds. Experimentally, 
Li$_x$FePO$_4$ compositions ($0 \leq x \leq 1$) are known to be unstable and phase 
separate into LiFePO$_4$ and FePO$_4$. However, first-principles calculations with 
LDA/GGA yield energetically favorable intermediate compounds and hence no phase 
separation. This qualitative failure of LDA/GGA seems to have its origin in the 
LDA/GGA self-interaction which delocalizes charge over the mixed-valence Fe ions, 
and is corrected by explicitly considering correlation effects in this material. 
This is demonstrated with LDA+U calculations which correctly predict phase separation 
in Li$_x$FePO$_4$ for $U-J \gtrsim 3.5$eV. The origin of the destabilization of 
intermediate compounds is identified as electron localization and charge ordering at 
different iron sites. Introduction of correlation also yields more accurate 
electrochemical reaction energies between FePO$_4$/Li$_x$FePO$_4$ and Li/Li$^+$ electrodes.
\end{abstract}

%
%
\maketitle
First-principles calculations employing density functional theory (DFT) 
have proven to be a powerful  method in understanding the thermodynamic, structural and electronic properties of a large class of materials.
The density functional is not known exactly, and is usually modeled within the Local Density Approximation (LDA) or Generalized Gradient Approximation (GGA). 
For many systems LDA or GGA gives remarkably good agreement with experiments, which has made these techniques valuable  tools to predict the behavior of materials \cite{ldagga}.
However, the self-interaction in LDA/GGA tends to delocalize electrons too much, and as such these methods are
unable to capture precisely the Coulomb correlation effects in correlated electron systems like transition metal oxides. 
The resulting failure to predict many  transition metal oxides as insulators has been well documented \cite{anisimovbook2000}.
%
%
%
In this paper we show by means of olivine-type \olix\ that the tendency for LDA/GGA to delocalize the $d$-electrons in mixed-valence transition metal oxides also leads to a qualitative failure in predicting miscibility and phase stability by a surprisingly large magnitude. The role Coulombic correlations play in phase stability  will be qualitatively probed.



\oli, a naturally occurring mineral, has attracted much attention recently, as its superb thermal safety, non-toxicity and low cost make it the most likely candidate for rechargeable Li-batteries electrodes in large applications such as electric and hybrid vehicles \cite{padhi1, padhi2, andersson, yamada-olivine, huang, prosini}.
In a battery, lithium is electrochemically and reversibly cycled in and out of the \oli\ material. As a result, the pseudo-binary 
FePO$_4$ - \oli\ phase diagram, critical for the material's behavior as an electrode, has been well characterized experimentally.

\begin{figure}
\includegraphics[width=0.8 \linewidth]{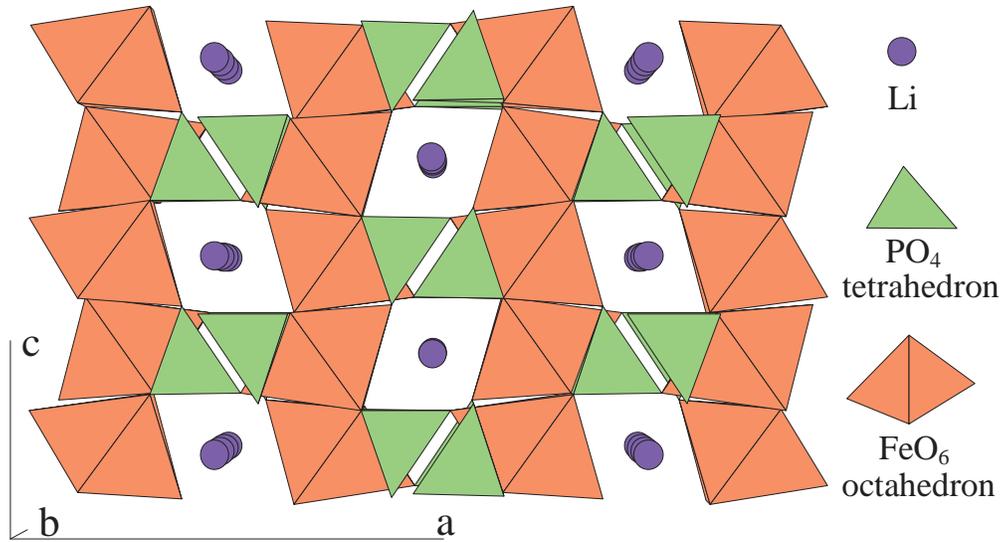}
\caption{Structure of \oli with cation polyhedra.
\label{fig:structure}}
\end{figure}

Olivine-type \oli\ and the de-lithiated structure \fepo\ have an orthorhombic unit cell with four Formula Units (FU) and space group Pnma (see Fig.\ \ref{fig:structure}). 
The olivine-type structure contains a distorted hexagonal close-packing of oxygen anions, with three types of cations occupying the interstitial sites: 1) corner-sharing FeO$_6$ octahedra which are nearly coplanar to form a distorted 2-d square lattice perpendicular to the {\bf a} axis, 2) edge-sharing LiO$_6$ octahedra aligned in parallel chains along the {\bf b} axis, and 3) tetrahedral PO$_4$ groups connecting neighboring planes or arrays. 
Electrochemical experiments and X-ray diffraction measurements have confirmed that no intermediate compound \olix\ exists between  \fepo\ and \oli\ \cite{padhi1,padhi2}, so that its phase diagram consists of a wide two-phase region with limited solubility on both the \fepo\ and \oli\ sides.
The magnetic structure of \oli\  and \fepo\ was determined from neutron diffraction data \cite{santoro-AFM, rousse-AFM}. Below the N\`eel temperature $T_N=50$K \cite{santoro-AFM} and 125K \cite{rousse-AFM}, respectively, the iron spins align in an antiferromagnetic (AFM) array, 
 induced by Fe-O-Fe superexchange interactions  between neighboring iron atoms. 

The objective of this paper is to investigate the stability of compounds between the composition \fepo\ and \oli\ and demonstrate that Coulomb correlations are essential in reproducing the absence of intermediate compounds. Different Li arrangements with 4 formula units are considered in the primitive cell. All possible symmetry-distinct decorations of the 4 Li sites gives seven structures, including two end members (x=0, 1), one structure at each of x=0.25 and 0.75, and 3 at x=0.5, here named 0.5a, 0.5b and 0.5c. The structures 0.5a, 0.5b and 0.5c have Li remaining at sites 1 and 2, 1 and 3, and 1 and 4, respectively (see Table \ref{tab:lithiumposition}). All the five intermediate structures have lower symmetry than the end members, and are monoclinic or triclinic.
\begin{table}[htbp]
\begin{ruledtabular}
\begin{tabular}{|c|c|c|c|c|c|c|c|c|}
	 & Li 1 & Li 2 & Li 3 & Li 4 & Fe 1 & Fe 2 & Fe 3 & Fe 4 \\ 
\hline x & 0 & 0.5 & 0.5 & 0 & .28 & .22 & .78 & .72 \\ 
\hline y & 0 & 0 & 0.5 & 0.5 & .25 & .75 & .25 & .75 \\ 
\hline z & 0 & 0.5 & 0.5 & 0 & .98 & .48 & .52 & .02 \\ 
\end{tabular} 
\end{ruledtabular}
\caption{Fractional positions of the four Li and four Fe atoms within the unit cell.
 \label{tab:lithiumposition}}
\end{table}
Total energy calculations were performed for the seven structures in GGA (or LDA where explicitly stated) with the projector-augmented-wave (PAW) method \cite{paw,paw-vasp} as implemented in  the Vienna Ab-initio Simulation Package \cite{vasp1}.  An energy cut-off of 500 eV and appropriate k-point mesh were chosen so that the total ground state energy is converged within 3meV per FU. All the atoms and cell parameters are fully relaxed at each structure. For x=0.25 and 0.75 the remaining $S_2$ point group symmetry has to be removed by imposing different initial magnetization on the irons to get the electronic ground state  (see below).
The results in this paper represent the ferromagnetic (FM) spin-polarized configurations unless stated explicitly. Although the magnetic ground state of \oli\ and \fepo\ is AFM \cite{santoro-AFM, rousse-AFM}, the difference
in FM and AFM formation energies (defined below) is a few meV/FU in most cases, not exceeding 12 meV,  and  does not affect the qualitative analysis, which is clearer in the
FM configuration. Iron is found to be always in the high-spin state, with the five majority spin $3d$-orbitals occupied.

Here we define $\Delta E(x)$, the formation energy  per FU of \olix\ as
\begin{equation}
\label{eq:formation-energy}
\Delta E(x)= E(x)-\left(x\, E\left(x=1\right) + \left(1-x\right) E\left(x=0\right)\right)
\end{equation} 
where $E(x)$ is the ground state total energy per FU for the structure with lithium concentration x. A negative formation energy means compound formation is energetically
favorable. In order for phase separation to occur at room temperature, all intermediate structures should have positive formation energy, large enough
to overcome the potential entropy gain in mixing. 
LDA results of $\Delta E(x)$ for all five structures are negative.
Although GGA slightly increases the formation energy, the prediction remains qualitatively in disagreement with experiment.
\begin{table}[htbp]
\begin{ruledtabular}
\begin{tabular}{|c|c|c|c|c|c|} 
	x    		& 0.25	& 0.5a	&  0.5b	& 0.5c	& 0.75 \\ 
\hline LDA 	& -155	& -255 	& -247 	& -136	& -168 \\ 
\hline GGA	& -135	& -209 	& -197 	& -129	& -138 \\ 
\end{tabular} 
\end{ruledtabular}
\caption{LDA and GGA formation energy (meV/FU) at different Li concentrations.
 \label{tab:ldagga}}
\end{table}

Given that the true formation energies should all be positive, these errors are large and somewhat surprising, since formation energies are properly weighted energy differences between similar structures, and as such usually benefit from significant error cancellations. For example, in many binary alloys formation energies are only 100$\sim$200meV/atom in magnitude, and hence large errors such as those found here would make them completely unreliable, which, based on the good agreement of many LDA/GGA studies with experiment,  is not the case \cite{alloy}. 

To investigate whether Coulombic on-site effects could be related to this substantial failure of LDA/GGA 
we carried out rotationally invariant \ldapu\ (GGA+$U$, more accurately) \cite{liechtenstein} calculations.
The essence of the method can be summarized by the expression for the total energy
\begin{equation} 
\label{eq:ldapu}
E_{{\rm LDA+}U}[\rho, \hat{n}]= E_{\rm LDA}[\rho] + E_{\rm Hub}[\hat{n}] - E_{\rm dc}[\hat{n}] 
	\equiv  E_{\rm LDA}[\rho] +E_{U}[\hat{n}] 
\end{equation} 
where $\rho$ denotes the charge density and $\hat{n}$ is the iron on-site $3d$ occupation matrix. The Hatree-Fock like interaction $E_{\rm Hub}$ from the Hubbard model replaces the double counting (dc) term $E_{\rm dc}$   representing the LDA on-site interaction. The $U$ correction term $E_{U}\equiv E_{\rm Hub} - E_{\rm dc}$ is defined by Eq. \ref{eq:ldapu}. However $E_{\rm dc}$ is not uniquely defined, and here we consider three common approaches \cite{dc}. The ``around mean field'' dc functional \cite{AMF} (dc1) yields low-spin iron, in disagreement with experiment \cite{rousse-AFM}. This is not surprising since dc1 usually works poorly in strongly correlated systems. We then compared formation energies with the dc functional defined in \cite{liechtenstein} (dc2) and with its spherically averaged version \cite{dudarev} (dc3). The latter reads
\begin{eqnarray}
E_{\rm dc}(\hat{n})  &=& \frac {U-J} 2  {\rm Tr}\hat{n}( {\rm Tr}\hat{n}-1)= \frac {U'} 2  {\rm Tr}\hat{n}( {\rm Tr}\hat{n}-1), \\  \label{eq:dc}
E_{U}(\hat n)&=& \frac {U-J} 2 {\rm Tr}\left(\hat{n} (1-\hat{n}) \right)= \frac {U'} 2 {\rm Tr}\left(\hat{n} (1-\hat{n}) \right),
\end{eqnarray}
where we have defined $U'=U-J$.
The formation energy with dc2 is very insensitive to a large range of $J$ (0- 2eV) when $U'$ is fixed, and agrees with  dc3 results within 10 meV for $U'\gtrsim $ 2eV. Therefore, we will use dc3, where there is only one effective parameter, $U'$. We evaluate all results as function of $U'$, spanning the range from 0 to 5.5eV. When calculating formation energies for a given $U'$, we assume $U'$ to be the same for all structures. The choice of $U'$ is a source of uncertainty in \ldapu\ calculations. However, we present the results as a function of $U'$ and will argue that the correct physics is obtained within a reasonable range of $U'$.


In Fig.\ \ref{fig:concentration} formation energies at  different $U'$ are shown as a function of Li concentration $x$. At each concentration $\Delta E$ increases with $U'$ and becomes positive at intermediate $U' (\approx$2.5-3.5eV). 
The formation energies saturate to a nearly constant value around $U'\approx $3.5-4.5eV.
The effect of the $E_U$ term is to drive the Fe-$3d$ orbital occupation numbers to integer (0 or 1) values.
As a result, the Fe ions tend to have integral occupancy even in the partially lithiated structures, and charge ordering occurs: we see distinct Fe$^{3+}$ and Fe$^{2+}$ in DFT+$U$ instead of the uniform Fe$^{(3-x)+}$ seen in LDA/GGA. 
For low $U'$ values ($U' \lesssim$ 1eV) the four Fe ions in the unit cell have similar $3d$ electron occupancy and Fe-O bond lengths for all the intermediate structures. Therefore, little charge ordering occurs in this limit, even though the Fe ions occupy symmetrically distinct positions. We will call these Fe cations (3-x)+ like. They are stable with respect to small perturbations in initial charge distribution.
In the high limit of $U'$($\gtrsim$3.5 or 4.5 eV)  there are 2 types of Fe 
ions, one very similar to those in FePO$_4$\ (which we call Fe$^{3+}$ like) and the other similar to those in \oli\ (called Fe$^{2+}$ like). The designation 3(2)+ is only meaningful in that the Fe ions are similar to those in \fepo(\oli). The Fe-O hybridization gives them less than nominal charge. 
For x=0.25(0.75) calculations imposing the symmetry of the structure on the charge density leads to two 3(2)+ like and two 2.5+ like Fe ions. Only when symmetry is broken does a lower energy state with three 3(2)+ like and one 2(3)+ like ions form. In these structures the charge density has lower symmetry than what would be expected from the ionic positions and, hence, charge ordering occurs.  
As the analysis  for all five structures is similar we choose x=0.5a as a typical intermediate structure for furthur discussion. 

\begin{figure}
\includegraphics[width=0.8 \linewidth]{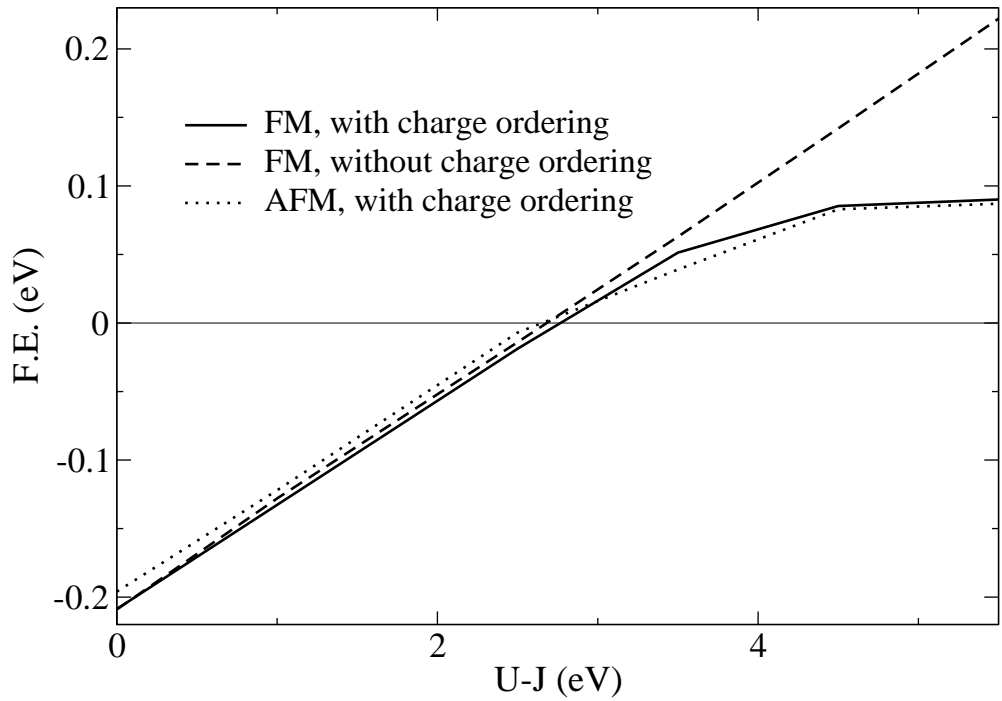}
\caption{Formation energy of structure 0.5a versus $U'$.
\label{fig:FE-U}}
\end{figure}

In Fig.\ \ref{fig:FE-U} $\Delta E$(x=0.5a) is shown as a function of $U'$.
We investigated AFM spin configurations in x=0, 0.5a and 1 and found them to give only slightly lower total energies. The AFM $\Delta E$ (dotted line) is almost equivalent to the FM one with charge ordering (solid line). 
We also studied a `restricted' FM system at x=0.5a where 
all four Fe ions have the same initial magnetization, ending up 2.5+ like. 
Charge ordering is absent in this metastable state, which has higher total energy than the charge-ordered ground state. 
From Fig.\ \ref{fig:FE-U} we can compare $\Delta E$ with and without charge ordering. 
Note that the curve with charge ordering levels off for $U\gtrsim 4.5$eV, which is explained below.

\begin{figure}
\includegraphics[width=0.8 \linewidth]{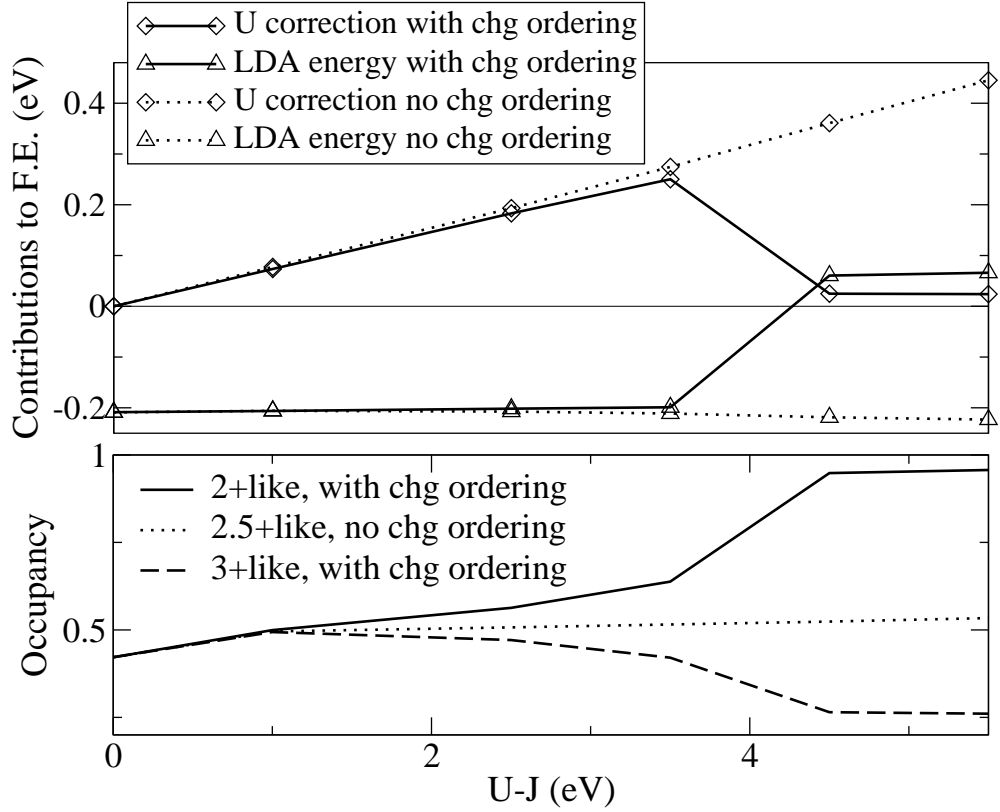}
\caption{Upper part: LDA (triangle) and $U$ correction term (diamond) contributions to $\Delta E$(x=0.5a)  {\it vs.} $U'$. Solid/dotted lines indicate presence/absence of charge ordering.
Lower part: occupancy of the most occupied minority-spin orbital versus $U'$, for Fe 2+ (solid line) and 3+ (dashed line) in the charge-ordered state and for 2.5+ (dotted line) in the state without charge ordering.
\label{fig:contrib-U}}
\end{figure}

To study quantitatively the change in formation energies and electron distribution as $U'$ is increased,  the contributions to $\Delta E$ are separated into the LDA energy,  $\Delta E_{\rm LDA}$,  and the  correction term, $\Delta E_U$, with definitions analogous to $\Delta E$ in Eq.\ \ref{eq:formation-energy}. The occupancy of the most occupied of the five minority-spin $3d$-orbitals of iron is displayed in the lower part of Fig.\ \ref{fig:contrib-U}. This orbital is most relevant  because its occupation makes the difference between Fe$^{3+}$ and Fe$^{2+}$.
When charge ordering is absent, the occupation number does not change much with $U'$ and stays near 0.5, as expected of a 2.5+ like Fe cation. In contrast, the curves in the charge-ordered state separate beyond $U'\approx 1$eV, with half of the ions becoming 2+ like and the other half 3+ like.
These occupancies can help to explain $\Delta E$ in the upper part of the diagram.
When charge ordering is absent (dotted lines) the four Fe cations in the x=0.5a structure are equally affected by $U'$ in terms of $3d$ occupation, as they are in x=0 and 1, and
the changes in ${\rm Tr}(\hat n (1-\hat n))$ in Li$_{0.5a}$FePO$_4$ are canceled by the weighted average of those in x=0 and 1 structures. 
As a result, the correction term  $\Delta E_U$ is almost proportional to $U'$, explaining its linear behavior in Fig.\ \ref{fig:contrib-U}.
When the symmetry is sufficiently broken, $\Delta E_U$ will make Fe-$3d$ charge density order so as to create, as much as possible, orbitals with integer occupation. This comes at a cost to $\Delta E_{\rm LDA}$, which changes from large negative values at $U'=0$ to positive values. We see two possible reasons why $\Delta E_{\rm LDA}$ increases when charge ordering occurs. Localization of the minority spin electrons into half of the Fe sites as Fe$^{2+}$ obviously leads to an increase in kinetic energy. Additionally, since Fe$^{2+}$ and Fe$^{3+}$ have different Fe-O bond lengths, their coexistence in one structure comes with a penalty in elastic energy. As the increase, relative to the LDA values, in $\Delta E_U$ is much smaller than in $\Delta E_{\rm LDA}$, the latter can be identified as the cause of phase separation.

The ground state electronic structure is also affected. The x=0.5a compound is insulating when charge ordering occurs in \ldapu, while it is metallic in LDA or \ldapu\ without charge ordering. The end members are  insulating in both LDA and \ldapu.

A weakness of the \ldapu\ method is that $U$ is an external parameter, and some justification for the choice of it is required. Considering a realistic $J=1eV$ \cite{dc} we find phase separation in the \olix\ system for $U\gtrsim 3.5$eV+$J$=4.5eV. Above this cutoff the formation energies and orbital occupancies become less sensitive to $U$.
The value of $U$  for these systems is likely to be even higher than this cutoff. A recent {\it ab-initio} computation of $U-J$ in the related Fe$_2$SiO$_4$ fayalite system suggests a value of 4.5eV for iron \cite{matteo}.  

\begin{figure}
\includegraphics[width=0.8 \linewidth]{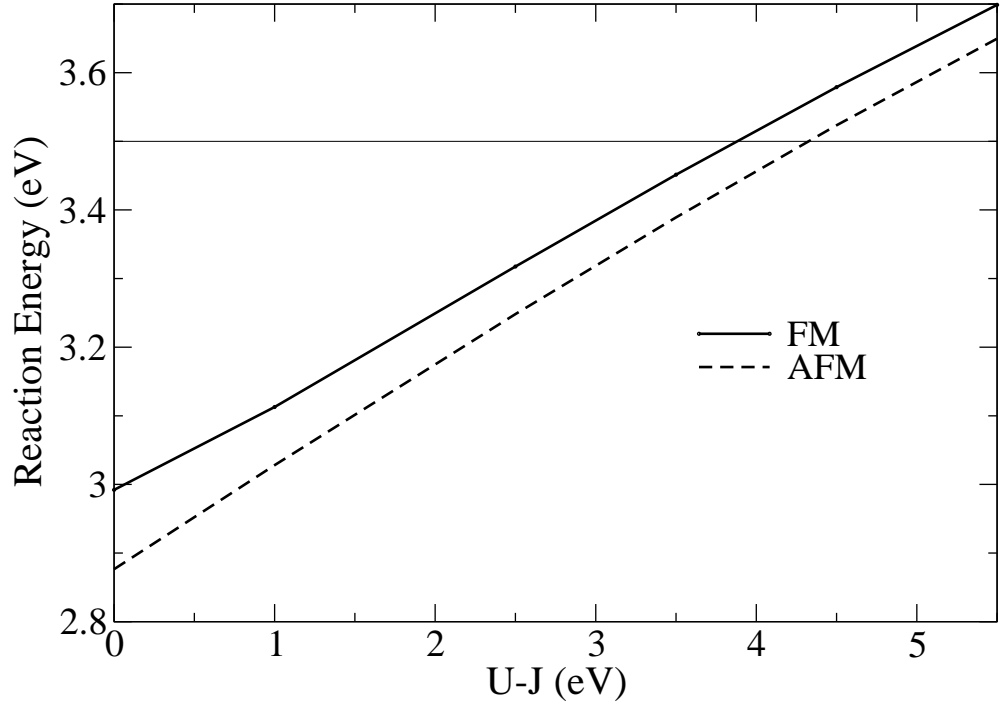}
\caption{Reaction energy in Eq. \ref{eq:battery-reaction} per FU versus $U'$ with FM
and AFM configurations, respectively.
\label{fig:voltage}}
\end{figure}

Another way to determine a physical value of $U$ is to compare the calculated and experimental reaction energy of \fepo\ and Li to form \oli.
\begin{equation}
\label{eq:battery-reaction}
E_{\rm reaction} = (E_{\rm Li}+ E_{\rm FePO_4} - E_{\rm LiFePO_4})  .
\end{equation} 
In this reaction Li$^+$ is inserted into the \fepo\ host and an electron is added to the $d$-states, reducing Fe$^{3+}$ to Fe$^{2+}$. Since the electron addition energy for Fe$^{3+}$ is a significant component of this reaction energy, the result will depend on the value of $U$. Experimentally, this energy can be measured very accurately, as it is the equilibrium electrical potential between \olix\ and Li-metal electrodes in a Li-electrolyte.
%
In Fig.\ \ref{fig:voltage} the calculated potential is plotted as a function of $U$ with FM and AFM spin configurations, respectively. The experimental voltage of 3.5V \cite{padhi2} is reached at $U-J \approx 4.2$eV. 

We have further confirmed that the positive formation energies obtained in Fig.\ \ref{fig:FE-U} are not an artifact of using a single unit cell by calculating the energy of four other structures (x=0.25 or 0.75) with a doubled unit cell. We found all these formation energies to be within $\pm$10meV of the results shown in Fig.\ \ref{fig:FE-U}.
Positive formation energies in GGA was recently confirmed in Ref. \cite{holzwarth}.

In summary, we find that both LDA  and GGA qualitatively fail to reproduce the experimentally observed phase stability and mixing energetics in the \olix\  system. 
For $U-J>3.5$eV, \ldapu\ calculations give positive $\Delta E$, in agreement with experiments. Hence, we speculate that the experimentally observed phase separation is due to the cost in kinetic and elastic energies when Fe$^{2+}$ and Fe$^{3+}$ coexist in \olix\ structures. This physics is not well captured by LDA/GGA, as the self-interaction causes a delocalization of the $d$-electrons, resulting in electronically identical Fe ions. As a result, there is no phase separation in LDA/GGA, in clear disagreement with experiment.

This work is supported by the Department of Energy under Grant DE-FG02-96ER45571 and by the National Science Foundation under Grant DMR-02-13282. F. Z. is grateful to Dr. E. Wu for his help in computation.


\bibliographystyle{prsty}


\begin{figure}
\includegraphics[width=0.8 \linewidth]{concentration.eps}
\caption{Formation energy of \olix \ at different x and $U'$ values. Points at x= 0.5 correspond to  structure 0.5a.
\label{fig:concentration}}
\end{figure}

\end{document}